\documentclass[useAMS,usenatbib,fleqn]{mnras}
\usepackage{newtxtext,newtxmath}
\usepackage{algorithm,algorithmic}
\usepackage[T1]{fontenc}
\usepackage{ae,aecompl}

\usepackage[dvips]{graphicx}
\usepackage{epsfig,amssymb,amsmath,color}
\newcommand{\beq}{\begin{equation}}
\newcommand{\eeq}{\end{equation}}
\newcommand{\beqn}{\begin{eqnarray}}
\newcommand{\eeqn}{\end{eqnarray}}
\DeclareMathOperator*{\argmin}{arg\,min}
\def\bmath#1{\mbox{\boldmath$#1$}}

\long\def\symbolfootnote[#1]#2{\begingroup%
\def\thefootnote{\fnsymbol{footnote}}\footnote[#1]{#2}\endgroup}

\usepackage{pstricks,pst-text}

\floatname{algorithm}{Algorithm}

\title[Data Multiplexing in Radio Interferometric Calibration]{Data Multiplexing in Radio Interferometric Calibration}
\author[Yatawatta et al.]{Sarod Yatawatta$^{1}$\thanks{E-mail:
yatawatta@astron.nl} Faruk Diblen$^{2}$ Hanno Spreeuw$^{2}$ L.V.E.~Koopmans$^{3}$\\
$^{1}$ASTRON, Postbus 2, 7990 AA Dwingeloo, The Netherlands\\
$^{2}$Netherlands eScience Center, Science Park 140,\\ 1098 XG Amsterdam, The Netherlands\\
$^{3}$Kapteyn Astronomical Institute, University of Groningen,\\ P.O. Box 800, 9700 AV Groningen, The Netherlands}

\begin{document}
\date{\today}
\pagerange{\pageref{firstpage}--\pageref{lastpage}} \pubyear{2017}
\maketitle
\label{firstpage}
\begin{abstract}
New and upcoming radio interferometers will produce unprecedented amounts of data that demand extremely  powerful computers for processing. This is a limiting factor due to the large computational power and energy costs involved. Such limitations restrict several key data processing steps in radio interferometry. { One such step is calibration} where systematic errors in the data are determined and corrected. Accurate calibration is an essential component in reaching many scientific goals in radio astronomy and the use of consensus optimization that exploits the continuity of systematic errors across frequency significantly improves calibration accuracy. In order to reach full consensus, data at all frequencies need to be calibrated simultaneously. { In the SKA regime, this can become intractable if the available compute agents do not have the resources to process data from all frequency channels simultaneously}. In this paper, we propose a multiplexing scheme that is based on the alternating direction method of multipliers (ADMM) with cyclic updates. With this scheme, it is possible to simultaneously calibrate the full dataset using far fewer compute agents than the number of frequencies at which data are available. 
We give simulation results to show the feasibility of the proposed multiplexing scheme in simultaneously calibrating a full dataset when a limited number of compute agents are available.
\end{abstract}
\section{Introduction}
In order to reach many scientific goals of modern radio astronomy, large amounts of interferometric data need to be collected to enable the detection of weak signals buried in the data. As a consequence, modern radio interferometers produce ever increasing amounts of data. { A case in point is the  Square Kilometre Array (SKA \cite{SKA0}) that is poised to surpass most existing radio interferometers in terms of data output.}
Interferometric data in raw form are affected by systematic errors such as the receiver beam and the Earth's atmosphere. These errors are determined and corrected during calibration. Calibration of a typical SKA type dataset is a heavily compute intensive task because the systematic errors vary with time, frequency and direction (position in the sky). The accuracy of calibration is paramount in the recovery of faint signals of scientific interest. In order to improve the accuracy, we can exploit the continuity of systematic errors across frequency. With the use of consensus optimization \citep{boyd2011,DCAL,Brossard2016} it has been shown that calibration can be improved and results based on real observations \citep{Patil2017} have already confirmed this. The systematic errors are modeled as polynomials in frequency and this model is used as a regularization term in calibration. In order to get the best benefit of consensus optimization, we need to simultaneously calibrate all the data available at all frequencies. { In a situation where the number of available compute agents is much less than the number of frequencies at which data are available, this may become problematic.} Similar problems have been encountered in radio interferometric imaging  where the simultaneous use of all available { data} is a daunting task for which multiple solutions have been proposed \citep{Meil2016,Degu2016,Onose2016,Onose2017}.

\begin{figure}
\begin{minipage}{0.98\linewidth}
\input{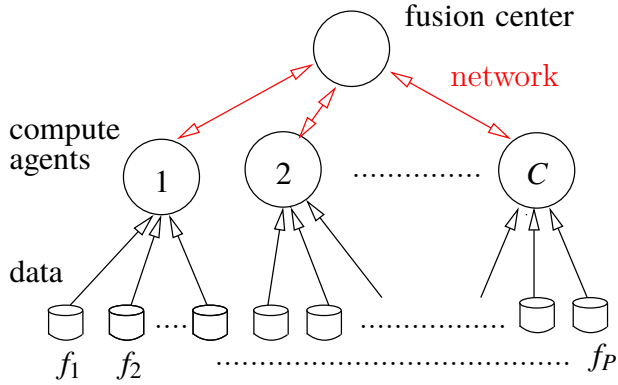}\\
\end{minipage}
\caption{Data are distributed across $C$ compute agents that are connected to the fusion center via a network. The total number of datasets ($P$) is larger than the number of available compute agents ($C$). Each dataset will have one or more frequency channels and the central frequencies $f_1,f_2,\ldots,f_P$ uniquely identify each dataset.\label{block}}
\end{figure}

We consider a situation where we have $P$ datasets distributed over $C$ compute agents as in Fig. \ref{block}. Each dataset will have data at one or more contiguous frequencies and we uniquely identify each dataset by its central frequency. We assume that each compute agent can process only a single dataset at a time, due to resource limitations (e.g. RAM\footnote{random access memory}, CPU\footnote{central processing unit}, GPU\footnote{graphics processing unit}).  Because of this assumption, in our previous work \citep{DCAL}, we restricted the number of datasets simultaneously processed to match the number of available compute agents. When $P \gg C$, this implies only processing $C$ datasets together (which we call as  {\em a comb}) to reach consensus  and repeating this until we have processed all $P$ datasets. However, using all $P$ datasets together to reach consensus is better than using multiple combs of size $C$.  One way to process all $P$ datasets together, when $P \gg C$, is {\em sequential} processing, where each compute agent uses its resources to sequentially process the data, and consensus is reached using all $P$ datasets. Sequential processing will be slower than processing a single comb, but will give better results. In this paper, we  try to achieve the improved result of sequential processing of all $P$ datasets, but with only spending the computational time required to process a single comb of $C$ datasets. In order to do this, we propose a data multiplexing scheme. 

Noting that calibration is a nonconvex optimization problem, we follow \cite{Hong15,Hong16} where the authors propose a {\em flexible} alternating direction method of multiplies (ADMM) algorithm. Only a subset of the available data is chosen for processing at each ADMM iteration, which is done in a cyclic manner (each slave cycles through the data one-by-one, also see section \ref{ssec:cadmm}). The crucial point in multiplexing is the selection of the penalty parameter, and as proposed by \citep{Hong15,Hong16}, the penalty needs to be as large as possible. However, unlike in other applications, consensus is achieved based on a model that does not represent the systematic errors with full accuracy. For instance, a simple phase screen will not represent ionospheric errors with sufficient accuracy beyond a certain threshold \citep{Martin2016}. The addition of beam errors \citep{Mort2016} confound this and systematic errors across a wide field of view cannot be guaranteed to be accurately represented by the consensus polynomials chosen in calibration. Therefore, selecting a penalty parameter that is too high will also give poor results due to the incomplete description of the systematic errors. Hence, we use an adaptive strategy to select the penalty parameter at each ADMM iteration. We base this on the Barzilai-Borwein method \citep{BB1988} as proposed by \cite{Zheng2016NIPS,Zheng2016,Zheng2017}.  The cyclic selection of different frequencies as in \citep{Hong16} and the adaptive update of the penalty parameter of each selected dataset as in \citep{Zheng2016NIPS} are combined in this paper. The initialization of the penalty is done using the Hessian of the cost function as proposed by \cite{EUSIPCO2016}. We  use the minimum description length (MDL) \citep{MDL} as a criterion for selecting  the consensus polynomials.

The rest of the paper is organized as follows: In section \ref{sec:calib}, we give an overview of direction dependent calibration using consensus optimization. In section \ref{sec:mux}, we introduce the data multiplexing scheme, starting with criteria for selecting the consensus polynomials (section \ref{ssec:mdl}), initialization of the penalty parameter (section \ref{ssec:init}), and adaptive update of the penalty (section \ref{ssec:update}). In section \ref{sec:simul}, we give results based on simulations of an SKA-like telescope (with $512$ stations) to demonstrate the performance of the proposed multiplexing scheme.

{\em Notation}: Upper case bold letters refer to matrices (e.g. ${\mathbfss{C}}$). Unless otherwise stated, all parameters are complex numbers. The set of complex numbers is given as ${\mathbb C}$ and the set of real numbers as  ${\mathbb R}$. The matrix pseudo-inverse, transpose, and Hermitian transpose are referred to as $(.)^{\dagger}$, $(.)^{T}$, $(.)^{H}$, respectively. The identity matrix (of size $N\times N$) is given by ${{\mathbfss {I}}_N}$. The Frobenius norm is given by $\|.\|$ and the cardinality of a set $\mathcal{F}$ is given by $|\mathcal{F}|$.

\section{Direction Dependent Radio Interferometric Calibration}\label{sec:calib}
We give a concise overview of direction dependent calibration with consensus optimization in this section. A more comprehensive overview is given in our previous work \citep{DCAL,EUSIPCO2016}.
\subsection{Data model}
We consider a radio interferometric array with $N$ dual-polarized receivers. The sky is composed of many discrete sources and we consider calibration along $K$ directions in the sky. The observed data at a baseline formed by two receivers, $p$ and $q$ ($\in[1,N]$), at frequency $f$ is given by \cite{HBS}
\beq \label{ME}
\mathbfss{V}({pqf})=\sum_{k=1}^K\mathbfss{J}_{pkf} \mathbfss{C}_{pqkf} \mathbfss{J}_{qkf}^H + \mathbfss{N}_{pqf}
\eeq
where $\mathbfss{V}({pqf})$ ($\in \mathbb{C}^{2\times 2}$) is the observed {\em visibility} matrix (or the cross correlations) at frequency $f$. The systematic errors that need to be calibrated for stations $p$ and $q$ are given by the Jones matrices $\mathbfss{J}_{pkf},\mathbfss{J}_{qkf}$ ($\in \mathbb{C}^{2\times 2}$), respectively. As calibration is performed along $K$ directions, each station has $K$ Jones matrices associated with it ($KN$ in total for the full array). The uncorrupted sky signal (or {\em coherency}) along the $k$-th  direction is given by $\mathbfss{C}_{pqkf}$ ($\in \mathbb{C}^{2\times 2}$) and is known a priori \citep{TMS}. The values of $\mathbfss{J}_{pkf},\mathbfss{J}_{qkf}$ and $\mathbfss{C}_{pqkf}$ in (\ref{ME}) are implicitly dependent on sampling time and frequency of the observation. However, their variation with $f$ is generally assumed to be smooth and enables the use of consensus optimization. The noise matrix $\mathbfss{N}_{pqf}$ ($\in \mathbb{C}^{2\times 2}$) is assumed to have complex, zero mean, circular Gaussian elements.

We use the space alternating generalized expectation maximization (SAGE) algorithm \citep{Fess94,Kaz2} to estimate $\mathbfss{J}_{pkf}$ for all possible values of $p$ and $k$ in (\ref{ME}). This reduces the computational cost and also improves the accuracy \citep{Kaz2}. Calibration along the $k$-th direction is done by using the effective observed data along the $k$-th direction
\beq \label{ME1}
\mathbfss{V}^k_{pqf} = \mathbfss{V}({pqf}) - \sum_{l=1,l\ne k}^K\widehat{\mathbfss{J}}_{plf} \mathbfss{C}_{pqlf} \widehat{\mathbfss{J}}_{qlf}^H
\eeq 
which is calculated using current estimates $\widehat{\mathbfss{J}}_{plf}$ and $\widehat{\mathbfss{J}}_{qlf}$. This is in fact the {\em expectation} step of the SAGE algorithm. The {\em maximization} step of the SAGE algorithm involves minimizing the objective function for the $k$-th direction defined under a Gaussian noise model as
\beq \label{cost1}
g_{kf}({\mathbfss{J}}_{1kf},{\mathbfss{J}}_{2kf},\ldots)= \sum_{p,q}\| {\mathbfss{V}}^k_{pqf} - {\mathbfss{J}}_{pkf} {\mathbfss{C}}_{pqkf} {\mathbfss{J}}_{qkf}^H \|^2.
\eeq
The summation in (\ref{cost1}) is over all the baselines $pq$ that are included in a finite time and frequency interval within which the systematic errors are estimated. It is also possible to modify the objective function for non-Gaussian noise models as in  \citep{Kaz3,SIRP,grobler2014}.

By using the SAGE algorithm, we are able to separate calibration along $K$ directions to $K$ independent calibration problems. Therefore, for the sake of simplicity, we drop $k$ from here onwards and rewrite the objective function (\ref{cost1}) for any general direction as
\beq \label{cost2}
g_{f}({\mathbfss{J}_f})= \sum_{p,q}\| {\mathbfss{V}}_{pqf} - {\mathbfss{A}}_p{\mathbfss{J}_f} {\mathbfss{C}}_{pqf} ({\mathbfss{A}}_q{\mathbfss{J}_f})^H \|^2
\eeq
where  ${\mathbfss{J}_f}$ ($\in \mathbb{C}^{2N\times 2}$) is the augmented matrix of Jones matrices for all stations along the $k$-th direction
\beq
{\mathbfss{J}_f}\buildrel\triangle\over=[{\mathbfss{J}}_{1kf}^T,{\mathbfss{J}}_{2kf}^T,\ldots,{\mathbfss{J}}_{Nkf}^T]^T,
\eeq
and ${\mathbfss{A}}_p$ ($\in \mathbb{R}^{2\times 2N}$) (and ${\mathbfss{A}}_q$ likewise) is the canonical selection matrix
\beq \label{Ap}
{\mathbfss{A}}_p \buildrel\triangle\over=[{\mathbfss{0}},{\mathbfss{0}},\ldots,{\mathbfss{I}_2},\ldots,{\mathbfss{0}}].
\eeq
Note that only the $p$-th block of (\ref{Ap}) is an identity matrix. To sum up, in direction dependent calibration along $K$ directions using data at frequency $f$, we solve $K$ subproblems of minimizing (\ref{cost2}).   

\subsection{Consensus optimization}
Consider $P$ datasets distributed across a network of $C$ compute agents as in Fig. \ref{block}. Rather than calibrating each dataset individually, we use consensus optimization to exploit the continuity of $\mathbfss{J}_f$ with $f$ to get an improved result. As introduced in \citep{DCAL,EUSIPCO2016}, we use ADMM algorithm for consensus optimization. We construct the augmented Lagrangian at the $n$-th ADMM iteration as
\beqn \label{aug}
\lefteqn{L_f({\mathbfss{J}}_f^n,{\mathbfss{Z}}^n,{\mathbfss{Y}}^n_f,\rho_f^n)}\\\nonumber
&&=g_{f}({\mathbfss{J}}_f^n) + \|\left(\mathbfss{Y}_f^n\right)^H({\mathbfss{J}}_f^n-{\mathbfss{B}}_f {\mathbfss{Z}^n})\| + \frac{\rho_f^n}{2} \|{\mathbfss{J}}_f^n-{\mathbfss{B}}_f {\mathbfss{Z}^n}\|^2
\eeqn
where we have used the superscript $(\cdot)^n$ to indicate the ADMM iteration number ($n=1,2,\ldots$) and the subscript $(\cdot)_f$ denotes data (and parameters) at frequency $f$. The original cost function $g_{f}({\mathbfss{J}}_f^n)$ is given by (\ref{cost2}). The Lagrange multiplier is given by ${\mathbfss{Y}}_f^n$ ($\in \mathbb{C}^{2N\times 2}$). The continuity in frequency is enforced by the frequency model given by ${\mathbfss{B}}_f$ ($\in \mathbb{R}^{2N\times 2NF}$), which is essentially a set of basis functions in frequency, evaluated at $f$. The number of terms used in each basis function is given by $F$. The global variable ${\mathbfss{Z}^n}$ ($\in \mathbb{C}^{2NF\times 2}$) is shared by data at all $P$ frequencies. The essential difference in (\ref{aug}) from our previous work is that we have the penalty $\rho^n_f$ to be variable both with $n$ and $f$.

An ADMM iteration for solving (\ref{aug}) is composed of three steps:
\beqn \label{step1}
({\mathbfss{J}}_f)^{n+1}= \underset{{\mathbfss{J}}}{\argmin}\ \ L_f({\mathbfss{J}},({\mathbfss{Z}})^n,({\mathbfss{Y}}_f)^n,\rho_f^n)\\ \label{step2}
({\mathbfss{Z}})^{n+1}= \underset{{\mathbfss{Z}}}{\argmin}\ \ \sum_f L_f(({\mathbfss{J}}_f)^{n+1},{\mathbfss{Z}},({\mathbfss{Y}}_f)^n,\rho_f^n)\\ \label{step3}
({\mathbfss{Y}}_f)^{n+1}=({\mathbfss{Y}}_f)^n + \rho^n_f\left( ({\mathbfss{J}}_f)^{n+1}-{\mathbfss{B}}_f ({\mathbfss{Z}})^{n+1} \right)
\eeqn
that are executed in sequence.  Summation across all frequencies at which data are available is denoted by $\sum_f$. The steps (\ref{step1}) and (\ref{step3}) are done for each $f$ in parallel. The update of the global variable (\ref{step2}) is done at the fusion center. More details of these steps can be found in \cite{DCAL}. In addition to the steps (\ref{step1},\ref{step2},\ref{step3}), we also extend \citep{CAMSAP2017} to update $\rho^n_f$, which is described in section \ref{ssec:update}.

The ADMM iterations (\ref{step1},\ref{step2},\ref{step3}) are initialized as follows:
\begin{itemize}
\item Normally ${\mathbfss{J}_f^1}$ is initialized using solutions obtained for the previous time interval, or using blocks of identity matrices ${\mathbfss{I}_2}$. 
\item The Lagrange multiplier ${\mathbfss{Y}}^1_f$ is set to ${\mathbfss{0}}$. 
\item Since ${\mathbfss{Z}}$ is estimated in closed form, no initialization is necessary. 
\item We will discuss the initialization of the penalty $\rho^1_f$ in section \ref{ssec:init}.
\end{itemize}

Even though all $C$ compute agents calibrate data in parallel, we assume that due to compute resource limitations, only one problem of type (\ref{step1}) can be solved at any time. We consider $P\gg C$ and introduce a multiplexing scheme in section \ref{ssec:cadmm} to handle this situation.

\section{Data multiplexing}\label{sec:mux}
We assume that the $P$ datasets are (approximately) evenly divided among the $C$ compute agents, in no particular order. { The} key point of the data multiplexing scheme is to achieve consensus (\ref{step2}) using all $P$ datasets, regardless of the value of $C$. We describe various aspects of the proposed multiplexing scheme in the following text.

\subsection{Selection of the consensus polynomial model}\label{ssec:mdl}
The consensus  polynomial functions used to construct ${\mathbfss{B}}_f$ in (\ref{aug}) are determined in advance. Given a choice of different polynomials, in particular with a varying number of terms $F$, we can use the minimum description length (MDL) \citep{MDL} to select the best polynomial model to use. Let one possible polynomial configuration (with $\widetilde{F}$ number of terms) construct $\widetilde{\mathbfss{B}}_f=\widetilde{\mathbfss{b}}_f^T \otimes \mathbfss{I}_{2N}$ ($\in \mathbb{R}^{2N\times 2\widetilde{F}N}$) at frequency $f$ using $\widetilde{\mathbfss{b}}_f \in\mathbb{R}^{\widetilde{F}\times 1}$ that represent the values of the $\widetilde{F}$ basis functions evaluated at $f$. Let the current solution be $\widetilde{\mathbfss{J}}_f$, which can be the solution after the first ADMM iteration (which is essentially the solution without consensus). We calculate the residual sum of squares (RSS) for this solution as
\beq \label{RSS}
\mathrm{{RSS}}= \frac{1}{8N} \sum_f \rho_f \| \widetilde{\mathbfss{J}}_f - \widetilde{\mathbfss{B}}_f {\mathbfss{Z}} \|^2 
\eeq
where we have calculated the RSS per parameter (because $\widetilde{\mathbfss{J}}_f \in \mathbb{C}^{2N\times 2}$ thus includes $8N$ real parameters). Using the RSS, we find the MDL as
\beq \label{MDL}
\mathrm{{MDL}}=\frac{P}{2} \log\left(\frac{\mathrm{RSS}}{P}\right) + \frac{\tilde{F}}{2} \log\left(P\right)
\eeq
and select the consensus polynomials (in particular $\widetilde{F}$) that give the minimum of (\ref{MDL}).

\subsection{Initialization of the penalty parameter}\label{ssec:init}
The original cost function (\ref{cost2}) is nonconvex and therefore its Hessian matrix is not positive definite. However, by carefully selecting the penalty parameter $\rho_f$, we might be able to make the Hessian matrix of the augmented Lagrangian (\ref{step1}) positive definite. Furthermore, the convergence of ADMM also depends on  $\rho_f$ \citep{Hong15,Hong16}.  The construction of the full Hessian matrix is computationally prohibitive. Therefore, we use the Hessian operator of the cost function (\ref{cost2}), which is given by \citep{DCAL,ICASSP13} as,
\beqn\label{Hess}
\lefteqn{\mathrm{Hess}_f\left(g_{f}({\mathbfss{J}}),{\mathbfss{J}},{\bmath \eta}\right)}\\\nonumber
&=&\sum_{p,q}\left( {\mathbfss{A}}_p^T \left( ({\mathbfss{V}}_{pqf}-{\mathbfss{A}}_p{\mathbfss{J}}{\mathbfss{C}}_{pqf}{\mathbfss{J}}^H{\mathbfss{A}}_q^T) {\mathbfss{A}}_q {\bmath \eta}\right.\right.\\\nonumber
&& \left.\left.- {\mathbfss{A}}_p({\mathbfss{J}}{\mathbfss{C}}_{pqf} {\bmath \eta}^H + {\bmath \eta}{\mathbfss{C}}_{pqf}{\mathbfss{J}}^H) {\mathbfss{A}}_q^T{\mathbfss{A}}_q{\mathbfss{J}}\right) {\mathbfss{C}}_{pqf}^H\right. \\\nonumber
&&\left. + {\mathbfss{A}}_q^T \left( ({\mathbfss{V}}_{pqf}-{\mathbfss{A}}_p{\mathbfss{J}}{\mathbfss{C}}_{pqf}{\mathbfss{J}}^H{\mathbfss{A}}_q^T)^H {\mathbfss{A}}_p {\bmath \eta}\right.\right.\\\nonumber
&& \left.\left.- {\mathbfss{A}}_q({\mathbfss{J}}{\mathbfss{C}}_{pqf} {\bmath \eta}^H + {\bmath \eta}{\mathbfss{C}}_{pqf}{\mathbfss{J}}^H)^H {\mathbfss{A}}_p^T{\mathbfss{A}}_p{\mathbfss{J}}\right) {\mathbfss{C}}_{pqf}\right) \\\nonumber
\eeqn
where  ${\bmath \eta}\in \mathbb{C}^{2N\times 2}$, $\mathrm{Hess}_f\left(g_{f}({\mathbfss{J}}),{\mathbfss{J}},{\bmath \eta}\right)\in \mathbb{C}^{2N\times 2}$. Note that ${\bmath \eta}$ is a matrix that { spans} the tangent space of the manifold (on which the minima of $g_{f}({\mathbfss{J}})$ lie) at ${\mathbfss{J}}$.

To investigate the positive definiteness of the Hessian, we need to find the smallest eigenvalue of (\ref{Hess}). Since we have a nonconvex cost function, the smallest eigenvalue is negative. As there is no closed form solution for the smallest eigenvalue, we use an iterative approach. First, we define a cost function $h({\bmath \eta})$ as \citep{EUSIPCO2016}
\beqn \label{hcost}
\lefteqn{h({\bmath \eta})\buildrel\triangle\over= \frac{1}{2}\mathrm{trace}\left({\bmath \eta}^H \mathrm{Hess}_f\left(g_{f}({\mathbfss {J}}),{\mathbfss {J}},{\bmath \eta}\right)\right.}\\\nonumber
&&+\left.\mathrm{Hess}_f^H\left(g_{f}({\mathbfss {J}}),{\mathbfss {J}},{\bmath \eta}\right) {\bmath \eta}\right)
\eeqn
and we  find the smallest eigenvalue $\lambda$ by solving
\beqn \label{eig}
&&\lambda=\underset{{\bmath \eta}}{\mathrm{min}}\ \ \ \ h({\bmath \eta})\\\nonumber
&&{\mathrm{subject\ to}}\ \ {\bmath \eta}^H{\bmath \eta}={\mathbfss {I}}_2.
\eeqn

There are several ways to solve (\ref{eig}). In our case, noting that the constraint ${\bmath \eta}^H {\bmath \eta}={\mathbfss{I}}_2$ makes the minimization of (\ref{hcost}) restricted onto a complex Stiefel manifold \citep{AMS}, we use the Riemannian trust region method \citep{RTR,manopt}. The gradient and Hessian of $h({\bmath \eta})$ are required to perform this optimization and are given as
\beq 
\mathrm{grad}\left(h({\bmath \eta}),{\bmath \eta}\right)= \mathrm{Hess}_f\left(g_{f}({\mathbfss{J}}),{\mathbfss{J}},{\bmath \eta}\right)
\eeq
 and 
\beq
\mathrm{Hess}\left(h({\bmath \eta}),{\bmath \eta},{\bmath \zeta}\right)=  \mathrm{Hess}_f\left(g_{f}({\mathbfss{J}}),{\mathbfss{J}},{\bmath \zeta}\right),
\eeq
where ${\bmath \zeta}$ ($\in \mathbb{C}^{2N\times 2}$) is a matrix that { spans} the tangent space  at ${\bmath \eta}$ of the Stiefel manifold. 

Note that the calibration solutions, i.e. ${\mathbfss J}$ are kept constant in (\ref{hcost}). We use the estimated solutions with $\rho_f$ set to zero and set this as ${\mathbfss J}$ in (\ref{hcost}). Once we have found the smallest eigenvalue, i.e. $\lambda$, we use this as a guideline to select $\rho_f$. The Hessian of the augmented Lagrangian (\ref{aug}) is given as 
\beq
\mathrm{Hess}_f\left( L_f({\mathbfss{J}},{\mathbfss{Z}},{\mathbfss{Y}}_f,\rho_f),{\mathbfss{J}},{\bmath \eta} \right) = \mathrm{Hess}_f\left(g_{f}({\mathbfss{J}}),{\mathbfss{J}},{\bmath \eta}\right) + \frac{\rho_f}{2} {\bmath \eta}
\eeq
where ${\bmath \eta}\in \mathbb{C}^{2N\times 2}$ has a similar definition as in (\ref{Hess}). So the Hessian of (\ref{step1}) has the smallest eigenvalue $\lambda+\rho_f/2$ where $\lambda$ is the smallest eigenvalue of (\ref{Hess}). By increasing $\rho_f > 2 |\lambda|$, we can make the minimization (\ref{step1}) convex. However, this also means that the penalty  is given more precedence than the actual cost function (\ref{cost2}). If the consensus polynomial chosen does not represent the systematic errors entirely accurately, increasing $\rho_f$ larger than $2 |\lambda|$ will lead to degraded performance as shown by \cite{EUSIPCO2016}. Therefore, we initialize $\rho_f$ to a fraction of $|\lambda|$, e.g., $\rho_f^{1}=|\lambda|/10$ and use $|\lambda|$ as an upper limit for $\rho_f$ in the adaptive update of $\rho_f$ as described in section \ref{ssec:update}.

The aforementioned initialization of $\rho_f$ is described for one direction out of $K$ directions. Although it is possible to solve (\ref{eig}) for each direction individually, it is much easier to scale the initial value of $\rho_f$ obtained for one direction to match other directions. Consider a rescaling of the model flux in (\ref{cost2}), i.e., ${\mathbfss{C}}_{pqf}$ is rescaled to $\kappa {\mathbfss{C}}_{pqf}$, where $\kappa$ is a positive scale factor. In this case, the solutions ${\mathbfss{J}}_f$ in (\ref{cost2}) are scaled to $\frac{1}{\sqrt{\kappa}}{\mathbfss{J}}_f$. Consequently, the penalty term $\frac{\rho_f^n}{2} \|{\mathbfss{J}^n}_f-{\mathbfss{B}}_f {\mathbfss{Z}^n}\|^2$ in (\ref{aug}) is also scaled by $\frac{1}{\kappa}$. Therefore, to get back the same penalty, we need to rescale $\rho_f$ to $\kappa \rho_f$. In other words, the penalty is scaled linearly with the scaling of sky model flux. Now consider rescaling of data ${\mathbfss{V}}_{pqf}$  in (\ref{cost2}) to $\omega {\mathbfss{V}}_{pqf}$, where $\omega$ is a positive scale factor. In this case, the solutions ${\mathbfss{J}}_f$ in (\ref{cost2}) are scaled to $\sqrt{\omega}{\mathbfss{J}}_f$ and all terms (both the cost function and the penalty term) in (\ref{aug}) are scaled by $\omega$. Therefore, in this case, no adjustment of $\rho_f$ is necessary.

In summary, the initialization of $\rho_f$ for one selected direction (out of $K$) is done by first determining the smallest eigenvalue of the Hessian, and using the magnitude of this as the upper bound for $\rho_f$. For the other $K-1$ directions, the initial values of $\rho_f$ are chosen by linear scaling according to the sky model flux. To minimize the extra compute overhead, the determination of $\rho_f$ need only be performed once for many calibration runs and, where appropriate, can be rescaled to obtain penalty factors for other scenarios.

\subsection{Adaptively updating the penalty parameter}\label{ssec:update}
In order to increase the convergence rate of ADMM, we update the penalty parameter at each ADMM iteration. Recent work by \cite{Zheng2016,Zheng2017}  { has} shown that by using the Barzilai-Borwein \citep{BB1988} method, ADMM can be accelerated in most practical applications. In particular, for nonconvex cost functions, \cite{Zheng2016NIPS} have shown that this adaptive update gives better results. Motivated by this, we use the same strategy in calibration. We have also compared another popular method for the update of $\rho_f$, which is called residual balancing \citep{He2000}. However, we have found that  \citep{CAMSAP2017} residual balancing is not stable in our case (also found in recent work \citep{Wohlberg2017}). We update the penalty only if we are confident of the performance improvement of ADMM with the update. One way of controlling the update is to use the $|\lambda|$ obtained in (\ref{eig}) as an upper bound for the updated value of $\rho_f$. We refer to this upper bound as $\overline{\rho}$ in the following text.

The update of $\rho_f$ at the $n$-th ADMM iteration is done according to algorithm \ref{algBB}. Prior to this update, (\ref{step1}) should be done at each slave and (\ref{step2}) should be done at the fusion center. The update of $\rho_f$ is done after (\ref{step3}) at each slave. Additional variables used at each slave are $\widehat{\mathbfss{Y}}_f^0,\widehat{\mathbfss{Y}}_f,\widehat{\mathbfss{J}}_f^0 \in \mathbb{C}^{2N\times 2}$.

\begin{algorithm}
\caption{Spectral penalty update at the $n$-th ADMM iteration for data at frequency $f$}
\label{algBB}
\begin{algorithmic}[1]

\REQUIRE Steps (\ref{step1}), (\ref{step2}) and (\ref{step3}) have been performed to obtain $({\mathbfss{J}}_f)^{n+1}$. Local variables $\widehat{\mathbfss{Y}}_f^0,\widehat{\mathbfss{Y}}_f,\widehat{\mathbfss{J}}_f^0 \in \mathbb{C}^{2N\times 2}$ are needed for each $f$ (and evolve with ADMM iterations). Upper bound for penalty is  $\overline{\rho}$ and $\underline{\alpha}\in(0,1]$ is a threshold parameter.
\IF {$n=1$}
\STATE Initialize $\widehat{\mathbfss{Y}}_f^0 :={({\mathbfss{J}}_f)^{1}}$
\ENDIF
\IF {$n$ is an iteration where $\rho_f$ is updated} \label{notalways}
\STATE $(\widehat{\mathbfss{Y}}_f)^{n+1}:=({\mathbfss{Y}}_f)^n + \rho^n_f\left( ({\mathbfss{J}}_f)^{n+1}-{\mathbfss{B}}_f ({\mathbfss{Z}})^{n} \right)$ \label{step4}

\STATE $\Delta {\mathbfss{Y}}_f := (\widehat{\mathbfss{Y}}_f)^{n+1} - \widehat{\mathbfss{Y}}_f^0, 
\Delta {\mathbfss{J}}_f := ({\mathbfss{J}}_f)^{n+1} - \widehat{\mathbfss{J}}_f^0$ \label{delta}

\STATE $\delta_{11}:=\mathrm{trace}\left(\mathrm{Real}\left(\Delta{\mathbfss{Y}}^H_f \Delta{\mathbfss{Y}}_f\right)\right)$ \label{delta11}
\STATE $\delta_{12}:=\mathrm{trace}\left(\mathrm{Real}\left(\Delta{\mathbfss{Y}}^H_f \Delta{\mathbfss{J}}_f\right)\right)$ \label{delta12}
\STATE $\delta_{22}:=\mathrm{trace}\left(\mathrm{Real}\left(\Delta{\mathbfss{J}}^H_f \Delta{\mathbfss{J}}_f\right)\right)$ \label{delta22}

\STATE $\alpha:=\frac{\delta_{12}}{\sqrt{\delta_{11}\delta_{22}}},
\alpha_{SD}:=\frac{\delta_{11}}{\delta_{12}}, \mathrm{and}\
\alpha_{MG}:=\frac{\delta_{12}}{\delta_{22}}$

\STATE $\hat{\alpha}:=
\begin{cases}
\alpha_{MG} & \mathrm{if}\ 2\alpha_{MG}>\alpha_{SD}\\
\alpha_{SD}-\frac{\alpha_{MG}}{2} & \mathrm{otherwise}
\end{cases}$

\STATE $\rho_f^{n+1}:=
\begin{cases}
\hat{\alpha} & \mathrm{if}\ \hat{\alpha}\le \overline{\rho}\ \ \mathrm{and}\ \alpha \ge \underline{\alpha}\\
\rho_f^{n} & \mathrm{otherwise} 
\end{cases}$ \label{rhoupdate}

\STATE $\widehat{\mathbfss{Y}}_f^0 :=(\widehat{\mathbfss{Y}}_f)^{n+1}\ \mathrm{and}\ 
\widehat{\mathbfss{J}}_f^0 :=({\mathbfss{J}}_f)^{n+1}$ \label{next}

\ENDIF

\end{algorithmic}
\end{algorithm}

Some remarks about algorithm \ref{algBB} are as follows:
\begin{itemize}
\item Local variables used (that do not live through ADMM iterations) are: $\Delta {\mathbfss{Y}}_f,\Delta {\mathbfss{J}}_f \in \mathbb{C}^{2N\times 2}$, $\delta_{11},\delta_{12},\delta_{22},\alpha,\hat{\alpha},\alpha_{SD},\alpha_{MG} \in \mathbb{R}$. The subscripts  $SD$ and $MG$ denote {\em steepest descent} and {\em minimum gradient}, respectively \citep{Zhou2006}.
\item Line \ref{notalways}: The penalty update is not performed at each ADMM iteration, \cite{Zheng2017} suggest updating at $T$  periodic values of $n$, where $T \ge 2$.
\item Line \ref{step4}: this update is different from (\ref{step3}) because $({\mathbfss{Z}})^{n}$ is used in the former and $({\mathbfss{Z}})^{n+1}$ is used in the latter. Therefore, the fusion center needs to temporarily store the old value of ${\mathbfss{Z}}$ at each iteration.
\item Line \ref{rhoupdate}:  The threshold $\underline{\alpha}$ $\in (0,1]$ is used to ensure that the changes in the Lagrange multiplier $\Delta {\mathbfss{Y}}_f$ and solutions $\Delta {\mathbfss{J}}_f$ on line \ref{delta} are sufficiently correlated (or have a positive direction cosine). We use $\underline{\alpha}=0.2$ as in \citep{Zheng2017} and by increasing this value, we can restrict the chances of spurious updates.
\item Line \ref{next}: $\widehat{\mathbfss{Y}}_f^0$ and $\widehat{\mathbfss{J}}_f^0$ are updated for use during the next update of $\rho_f$.
\end{itemize}

The additional computational cost needed to perform the adaptive update of $\rho_f$ is mostly due to  three linear operations (lines \ref{step4}, \ref{delta}) and three inner products (lines \ref{delta11},\ref{delta12},\ref{delta22}), all involving matrices in $\mathbb{C}^{2N\times 2}$. In addition, there is increased network communication overhead because the updated values of $\rho_f$ have to be passed to the fusion center and also because of the additional update on line \ref{step4} where ${\mathbfss{B}}_f ({\mathbfss{Z}})^{n}$ has to be sent from the fusion center to each slave.

\subsection{Cyclic ADMM with data multiplexing}\label{ssec:cadmm}
We first introduce the concept of a cyclic buffer. Let $\mathcal{F}$ contain a set of real numbers (e.g. $\mathcal{F}=\{f_1,f_2,f_3\}$) that in our case correspond to a set of frequencies. Consider a function $\mathrm{First}(\mathcal{F})$: Every time $\mathrm{First}(\cdot)$ is applied on $\mathcal{F}$, it will return the first entry of $\mathcal{F}$ and move this first entry to the last position of $\mathcal{F}$. So if $\mathcal{F}=\{f_1,f_2,f_3\}$, repeatedly calling $\mathrm{First}(\cdot)$ on $\mathcal{F}$ will give us $f_1,f_2,f_3,f_1,f_2,\ldots$, in other words $\mathrm{First}(\mathcal{F})$ will give us the elements in $\mathcal{F}$ repeatedly, in a cyclic manner. 

We use a cyclic buffer to represent the data locally available to each slave, assuming each dataset is uniquely identified by its frequency (or central frequency if each dataset contains more than one channel). For example, if slave $i$ has data at frequencies $\{f_1,f_2,f_3\}$ locally available, we use $\mathcal{F}_i=\{f_1,f_2,f_3\}$ where $\mathcal{F}_i$ is a cyclic buffer. With the use of a cyclic buffer, we give the pseudocode for cyclic ADMM in algorithm \ref{algCADMM}.

\begin{algorithm}
\caption{Cyclic ADMM with data multiplexing}
\label{algCADMM}
\begin{algorithmic}[1]
\REQUIRE $\mathcal{F}_i$ ($\subset \{f_1,f_2,\ldots,f_P\}$) is a  cyclic buffer that represent the data being calibrated by slave $i$. $\mathcal{W}_i$ is a set of frequencies of the data being calibrated during one ADMM iteration by slave $i$. $M$ is the maximum number of ADMM iterations. $T (\ge 2)$ is an integer that determines the periodicity of the penalty parameter update.
\STATE Randomly permute $\mathcal{F}_i$ \label{random}
\FOR{$n=1,\ldots,M$}
 \IF{$n=1$ or $n=M$} 
   \STATE $\mathcal{W}_i := \mathcal{F}_i$  \label{selectall}
 \ELSE
   \STATE  $\mathcal{W}_i :=\mathrm{First}(\mathcal{F}_i)$ \label{selectone}
 \ENDIF
 \FOR{$i=1,\ldots,C$ in parallel}
   \STATE Perform (\ref{step1})  $\forall f \in \mathcal{W}_i$ 
 \ENDFOR
 \STATE Perform (\ref{step2}) at the fusion center \label{fusion}
 \FOR{$i=1,\ldots,C$ in parallel}
   \STATE Perform (\ref{step3}) $\forall f \in \mathcal{W}_i$ 
 \ENDFOR

 \FOR{$i=1,\ldots,C$ in parallel}
   \STATE \COMMENT{Decide whether to update the penalty or not}
   
   \STATE $do\_update:=0$ \COMMENT{Default is no update}
   \IF{$|\mathcal{F}_i| > 1$}
     \STATE $do\_update:=1$
   \ELSIF{$|\mathcal{F}_i| = 1$ and $n>1$ and  $n$ is a multiple of $T$}
     \STATE $do\_update:=1$
   \ENDIF
   \IF{$do\_update=1$}
    \STATE Perform algorithm \ref{algBB} to update $\rho_f$ $\forall f \in \mathcal{W}_i$ \label{rupdate}
   \ENDIF
 \ENDFOR
 
\ENDFOR
\end{algorithmic}
\end{algorithm}

Some remarks on algorithm  \ref{algCADMM} are as follows:
\begin{itemize}
\item Line \ref{random}: the order of selection of data is randomized for each calibration run. A typical observation will contain many time samples and for each time sample, the ordering of the  frequencies of the data calibrated by each slave is random.
\item Lines \ref{selectall}, \ref{selectone}: At the first and the last ADMM iterations, (\ref{step1}) and (\ref{step3}) are performed on all available data, possibly in a sequential manner. On the other hand, in all other ADMM iterations, only a single dataset is selected for performing these steps. Thus, depending on the ADMM iteration, $\mathcal{W}_i$ can point to all available datasets for slave $i$ or just one dataset. Therein lies the multiplexing of data, where in most ADMM iterations, each slave works on a single dataset.
\item Line \ref{fusion}: Step (\ref{step2}) is performed at the fusion center using all frequencies, regardless of whether the datasets are selected by the slaves for processing or not. The flexible ADMM algorithm presented in \citep{Hong16} does not necessarily perform (\ref{step2}) at each ADMM iteration. This in essence converts algorithm \ref{algCADMM} to a sequential processing version of ADMM with some delay. 
\item Line \ref{rupdate}: The penalty parameter update is only done on the data at frequencies in $\mathcal{W}_i$, so in most ADMM iterations, for just one dataset. Thus, the penalty update interval for any given dataset on slave $i$ will be about $|\mathcal{F}_i|$, which is generally larger than the period $T$.
\end{itemize}

The performance of algorithm  \ref{algCADMM} depends on the value of $C$, especially for solving (\ref{step2}). We investigate this dependence further by using simulations in section \ref{sec:simul}. 

\section{Simulations}\label{sec:simul}
We give results based on simulations of an SKA-like telescope (with $N=512$ stations) in this section. The configuration of the stations is similar to the one used by \cite{Mort2016} and the integration time is 10 sec. We simulate data at $P=24$ different frequencies, spread in the range $115-185$ MHz, but note that in real observations, $P$ could be several hundred or more. The sky consists of $K=10$ bright point sources, spread across a field of view of $7\times 7$ square degrees, with peak fluxes in the range $[1.5,10]$ Jy and another $6000$ weak sources (which are not known during calibration) with peak fluxes in the range $[0.01,0.1]$ Jy. Systematic errors along the $K$ directions are randomly generated, with continuity across frequency created by an $8$ order ordinary polynomial in frequency and slow variability across time. The $6000$ weak sources are clustered \citep{Kazemi3} around the bright $K$ sources and also corrupted by the systematic errors of each cluster that it belongs to using (\ref{ME}). Finally, complex circular white Gaussian noise is added to the simulated data with a signal to noise ratio (ratio of signal power versus noise power) of $10$. The consensus polynomial (Bernstein basis functions) has $F=3$ terms that is chosen according to section \ref{ssec:mdl}. Note that this is well below the order of the polynomial that it used to generate the errors.

During calibration, the $6000$ weak sources are not used in the sky model and thus they act as additional noise. The systematic errors along the $K$ directions are estimated at each of the $P$ frequencies, per each time sample of 10 sec duration.  In order to  measure the performance of calibration (in particular the convergence of ADMM), we use the error between the ground truth value of  $\mathbfss{J}_f$ (per direction and frequency) and its estimated value at the $n$-th ADMM iteration $\widehat{\mathbfss{J}}_f^n$, calculated as  $\frac{1}{\sqrt{4N}}\|{\mathbfss{J}}_f-\widehat{\mathbfss{J}}_f^n {\mathbfss{U}}\|$ with a proper unitary matrix ${\mathbfss{U}}$ ($\in {\mathbb C}^{2\times 2}$) \citep{interpolation}, and averaged over all $K$ directions and $P$ frequencies. Additionally, we also calculate the primal residual $\| \mathbfss{J}_f^n - {\mathbfss{B}}_f {\mathbfss{Z}}^n \|$ and the dual residual $\| \rho_f^{n}{\mathbfss{B}}_f ({\mathbfss{Z}}^n -{\mathbfss{Z}}^{n-1}) \|$ produced in (\ref{step1}),(\ref{step2}) and (\ref{step3}), which are also averaged over all directions and frequencies. { The primal residual represents the error between the systematic errors predicted by the global model and its local estimate at each frequency. The dual residual represents the convergence of the global model to a stable value.} 

We compare four calibration scenarios using the simulated data. In all cases, the initial values of the penalty parameter is the same and for a source with $1$ Jy peak flux, the initial value we use for the penalty is about $10$ (determined according to section \ref{ssec:init} using data at frequency $148$ MHz) and this value is scaled according to the flux for all $K$ sources as described in section \ref{ssec:init}. The four calibration scenarios are as follows:
\begin{enumerate}
\item [(i)] $C=24$: Calibration using $C=P=24$ compute agents (no multiplexing but adaptive update of penalty as in algorithm \ref{algBB} with $T=2$). 
\item [(ii)] $C=12$ (multiplexing): Calibration using $C=12$ compute agents (multiplexing as in algorithm \ref{algCADMM}).
\item [(iii)] $C=8$ (multiplexing): Calibration using $C=8$ compute agents (multiplexing as in algorithm \ref{algCADMM}).
\item [(iv)] $C=8$ (no multiplexing): Calibration using $C=8$ compute agents $P/C=3$ times, (no multiplexing but with adaptive penalty update as in algorithm \ref{algBB} with $T=2$) where each comb is made of randomly selected data at $C$ frequencies.
\end{enumerate}
 Note that scenario i with $C=24$ is also equivalent to sequential processing of the $P$ frequencies using fewer compute agents if $C<P$. Scenario iv is described in \citep{DCAL} (but without adaptive update of the penalty parameter) and this paper aims to improve \citep{DCAL} but without the expenditure of additional compute agents nor reverting to sequential processing.

\begin{figure}
\begin{minipage}{0.98\linewidth}
\centerline{\epsfig{figure=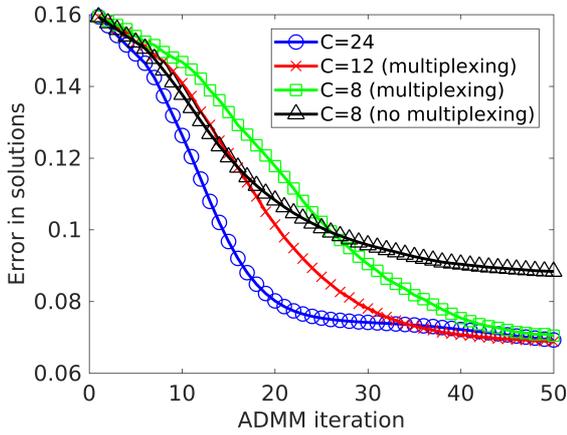,width=8.0cm}}
\end{minipage}
\caption{Variation of the error in solutions with ADMM iterations.\label{fig-solution-error}}
\end{figure}

The variation of the error in solutions with ADMM iteration $n$, compared to the ground truth value is shown in Fig. \ref{fig-solution-error}. We see that scenario i gives the fastest convergence and the lowest error. Multiplexing with $C=12$ and $C=8$ (scenarios ii and iii) gives increasingly slower convergence. On the other hand, scenario iv where $C=8$ and no multiplexing is done gives faster convergence in the beginning, but reaches a higher error floor. 

\begin{figure}
\begin{minipage}{0.98\linewidth}
\centerline{\epsfig{figure=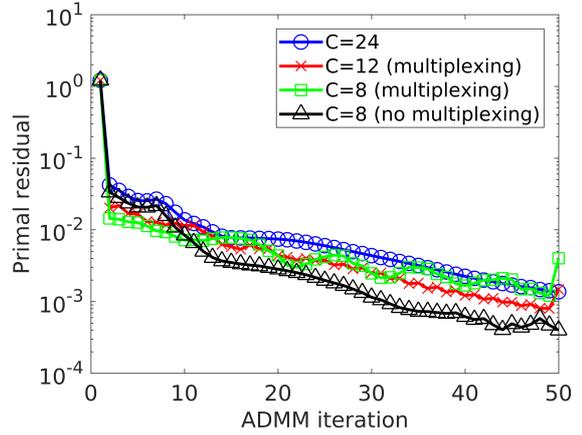,width=8.0cm}}
\end{minipage}
\caption{Variation of the primal residual with ADMM iterations.\label{fig-primal-residual}}
\end{figure}

\begin{figure}
\begin{minipage}{0.98\linewidth}
\centerline{\epsfig{figure=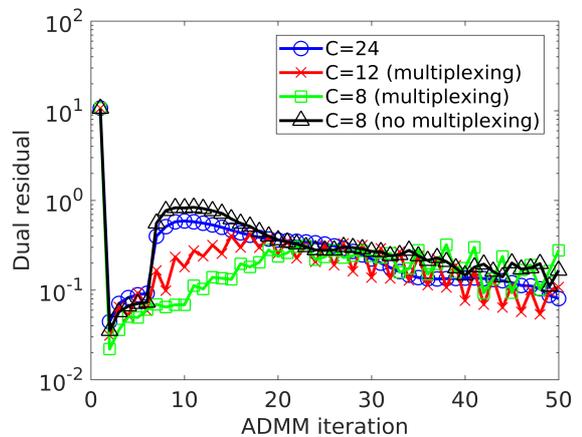,width=8.0cm}}
\end{minipage}
\caption{Variation of the dual residual with ADMM iterations.\label{fig-dual-residual}}
\end{figure}

The primal and dual residuals are shown in Figs. \ref{fig-primal-residual} and \ref{fig-dual-residual} respectively. Out of these two, the dual residual variation in Fig. \ref{fig-dual-residual} shows clear differences in the four calibration scenarios considered, that can also explain the different convergence rates seen in Fig. \ref{fig-solution-error}. First, we see that multiplexing leads to oscillatory behavior of the dual residual (and also of the primal residual to a lesser extent). This can be explained by the selection of subsets of frequencies for processing as in algorithm \ref{algCADMM} at each ADMM iteration, thus leading to a {\em limit cycle}. Secondly, in Fig. \ref{fig-dual-residual}, between $n\approx 8$ and $n \approx 20$ we see a marked difference in the dual residual between multiplexing (scenarios ii and iii) and no multiplexing (scenarios i and iv). Higher dual residuals mean faster convergence to the final value of ${\mathbfss{Z}}$ and we see that keeping the frequencies fixed to perform (\ref{step2}) enables faster convergence to the final value of ${\mathbfss{Z}}$. However, because the model ${\mathbfss{B}}_f$ is incomplete, the final value of ${\mathbfss{Z}}$ of each comb (scenario iv) can converge to values different than the final value at convergence using the full data (scenario i). This can also make the primal residual lower for each comb for calibration scenario iv, which is seen in Fig. \ref{fig-primal-residual}. This does not mean that the actual error in solutions is lower for scenario iv, as seen in Fig. \ref{fig-solution-error}. { At the last ADMM iteration, for scenarios ii and iii, we see an increase in the primal and dual residuals. This is because we use all available data to reach consensus at the last ADMM iteration (see lines 3,4,5 in algorithm \ref{algCADMM}). However, this does not increase the error in solutions as seen in Fig. \ref{fig-solution-error}.}

We draw several conclusions from Figs. \ref{fig-solution-error},\ref{fig-primal-residual} and \ref{fig-dual-residual}. First, calibration using all available data (scenario i), either by using more compute agents or by sequential processing, gives the best results. The convergence of data multiplexing (scenarios ii and iii) is slower, mainly because of the convergence of the global variable ${\mathbfss{Z}}$ is slow. However, we can still get the desired accuracy albeit with more ADMM iterations. The main advantage of scenarios ii and iii as opposed to scenario i is the computational cost: either because it uses fewer compute agents or because it requires less computations per each ADMM iteration. To elaborate, consider  the total cost required in scenario i compared to scenario iii: In scenario i, we reach the error floor in about $20$ ADMM iteration while in scenario iii, this is about $40$. However, scenario iii uses $1/3$ compute agents (or compute cycles in sequential processing). Therefore, the total cost of scenario iii is $40/3 \approx 14$, which is less than in scenario i. Secondly, there is clear improvement in processing all available data (with or without multiplexing as in scenarios i,ii and iii) compared to processing subsets (or combs) of data as in scenario iv. In other words, it is possible to improve scenario iv without expending more computations by multiplexing, but multiplexing will not necessarily give the best achievable result (scenario i). Moreover, the performance of multiplexing degrades as the amount of multiplexing increases, or as fewer compute agents are used to process the same amount of data. This can be seen from comparing the performance of scenario ii with scenario iii.

\begin{figure}
\begin{minipage}{0.98\linewidth}
\centerline{\epsfig{figure=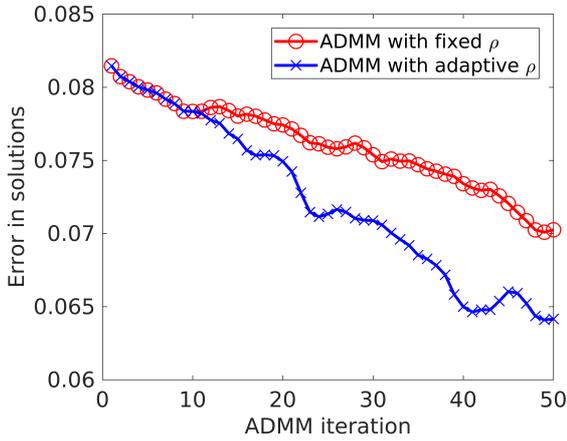,width=8.0cm}}
\end{minipage}
\caption{Variation of the error in solutions with ADMM iterations, for fixed penalty parameter and for adaptive penalty parameter.\label{solution-error-image}}
\end{figure}

\begin{figure}
\begin{minipage}{0.98\linewidth}
\centerline{\epsfig{figure=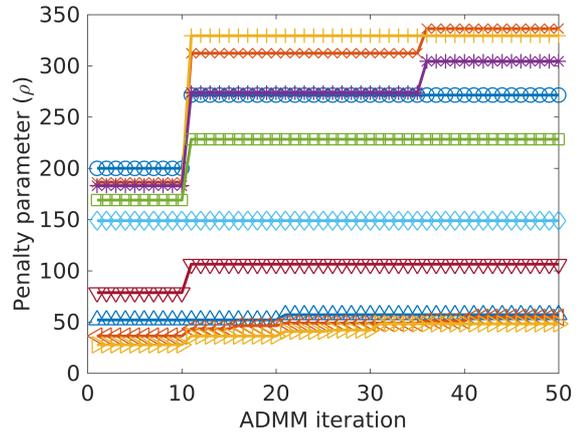,width=8.0cm}}
\end{minipage}
\caption{Variation of the penalty parameter $\rho$ with ADMM iterations for all $10$ directions at one frequency. The initial values of $\rho$ are scaled according to the flux of the source being calibrated as described in section \ref{ssec:init}. We see that $\rho$ update occurs at very few instances. Not all directions have updates of $\rho$ and the updates does not happen at the same ADMM iteration for all directions.\label{rho-admm-image}}
\end{figure}

The effect of the adaptive penalty update is subtle and we emphasize that the penalty is updated only if the new value for the penalty can be estimated with some confidence (that can be controlled by the threshold $\underline{\alpha}$ in (\ref{rhoupdate})). This is similar to the conclusions drawn by \cite{Zheng2016NIPS}. Therefore, in order to compare the effect of the adaptive update of the penalty,  we give a comparison of data multiplexing (scenario iii) with and without the adaptive penalty update (also see \citep{CAMSAP2017}).  We show the error in solutions in Fig. \ref{solution-error-image}, with fixed penalty and with adaptive update of the penalty. Moreover, we show the variation of the penalty parameter at one value of $f$ for this example in Fig. \ref{rho-admm-image}. Note that in Fig. \ref{rho-admm-image}, the initial value of the penalty is different for each of the $K$ directions (scaled according to the flux) and the variation of the penalty is also different for each direction. Nonetheless, as seen in Fig. \ref{solution-error-image}, the adaptive update of the penalty shows faster reduction in the error in solutions.

Due to the increased number of stations ($N=512$) and hence the amount of data, an important issue that needs clarification is the computational cost of calibration. As shown by \cite{Kaz2}, the scaling of consensus optimization with the number of directions being calibrated $K$ is linear, mainly due to the use of the SAGE algorithm. The scaling with the number of stations $N$ depends on the low level optimization routine used in consensus optimization. We use the Riemannian trust region algorithm \citep{RTR,ICASSP13} as the underlying optimization routine. In this algorithm, the linear optimization is done using the truncated conjugate gradient method \citep{RTR} with matrices in ${\mathbb{C}^{2N \times 2}}$, and therefore the direct solution of a linear system is not needed. This algorithm scales linearly with $N$ because the size of matrices in ${\mathbb{C}^{2N \times 2}}$ grows linearly with $N$. The dominating cost is mostly due to the model ${\mathbfss{C}}_{pqf}$ computation in (\ref{cost2}) as well as computing the cost function together with its gradient and the Hessian. This needs to be done for each data point and the number of data points scales as $N^2$ (baselines), and linearly with $K$ and $P$.

\section{Conclusions}\label{sec:conclusions}
In order to simultaneously process data at a large number of frequencies with a limited number of compute agents, we have proposed a multiplexing scheme for consensus optimization. Based on simulation results, we conclude that the multiplexing scheme together with the adaptive update of the penalty parameter improves the quality of direction dependent calibration when the compute resources are limited. The source code for the algorithms described in this paper is available at http://sagecal.sf.net/ and https://github.com/nlesc-dirac/sagecal where we have used the message passing interface (MPI) as our network communication framework. Future work will focus on migrating these algorithms to big-data frameworks such as Apache Spark.

\section*{Acknowledgments}
This work is supported by Netherlands eScience Center (project DIRAC, grant 27016G05) and the European Research Council (project LOFARCORE, grant 339743). We thank the anonymous reviewer and Ronald Nijboer for valuable comments.

\bibliographystyle{mnras}
\bibliography{references}
\bsp
\label{lastpage}
\end{document}